\pdfoutput=1

\documentclass[11pt]{article}

\usepackage[preprint]{acl}

\usepackage{times}
\usepackage{latexsym}

\usepackage[T1]{fontenc}

\usepackage[utf8]{inputenc}

\usepackage{microtype}
\usepackage{amsmath}
 \usepackage{booktabs}
 \usepackage{amssymb}
\usepackage{CJKutf8}
\usepackage{inconsolata}

\usepackage{graphicx}
\usepackage{stfloats}

\title{HyReC: Exploring Hybrid-based Retriever for Chinese}

  \author{Zunran Wang  \\\And
  Shenpeng Zheng \\\And
  Shenglan Wang \\
  Huawei Technologies Ltd. Co. \\\And
  Minghui Zhao \\\And
  Zhonghua Li \\
  }

\begin{document}
\maketitle
\begin{abstract}
Hybrid-based retrieval methods, which unify dense-vector and lexicon-based retrieval, have garnered considerable attention in the industry due to performance enhancement. However, despite their promising results, the application of these hybrid paradigms in Chinese retrieval contexts has remained largely underexplored. In this paper, we introduce HyReC, an innovative end-to-end optimization method tailored specifically for hybrid-based retrieval in Chinese. HyReC enhances performance by integrating the semantic union of terms into the representation model. Additionally, it features the Global-Local-Aware Encoder (GLAE) to promote consistent semantic sharing between lexicon-based and dense retrieval while minimizing the interference between them. To further refine alignment, we incorporate a Normalization Module (NM) that fosters mutual benefits between the retrieval approaches. Finally, we evaluate HyReC on the C-MTEB retrieval benchmark to demonstrate its effectiveness.
\end{abstract}

\section{Introduction}
Retrieval-augmented generation (RAG) enhances large language models by incorporating external knowledge to address hallucination issues, simultaneously catalyzing the rapidly evolving development of the retrieval community. How to effectively retrieve the most relevant information from the knowledge base is critically important for the final generation results. According to the encoding space, retrieval methods can be mainly categorized into three classifications: dense-vector( e.g.,   Condenser~\cite{gao2021condenser}, Bge embedding~\cite{bge_embedding}, and Jina embedding~\cite{sturua2024jinaembeddingsv3multilingualembeddingstask}), lexicon-based( e.g., DeepCT~\cite{dai2019context}, SparTerm~\cite{bai2020sparterm}, and TILDE~\cite{zhuang2021tilde})
and hybrid-based paradigms(e.g., COIL-full~\cite{gao2021coil}, Unifier~\cite{shen2023unifier} and Bge M3~\cite{bge-m3} ). Among them, the hybrid-based paradigm has garnered significant attention owing to its superior performance. 



\begin{figure}[t]
  \includegraphics[width=\columnwidth]{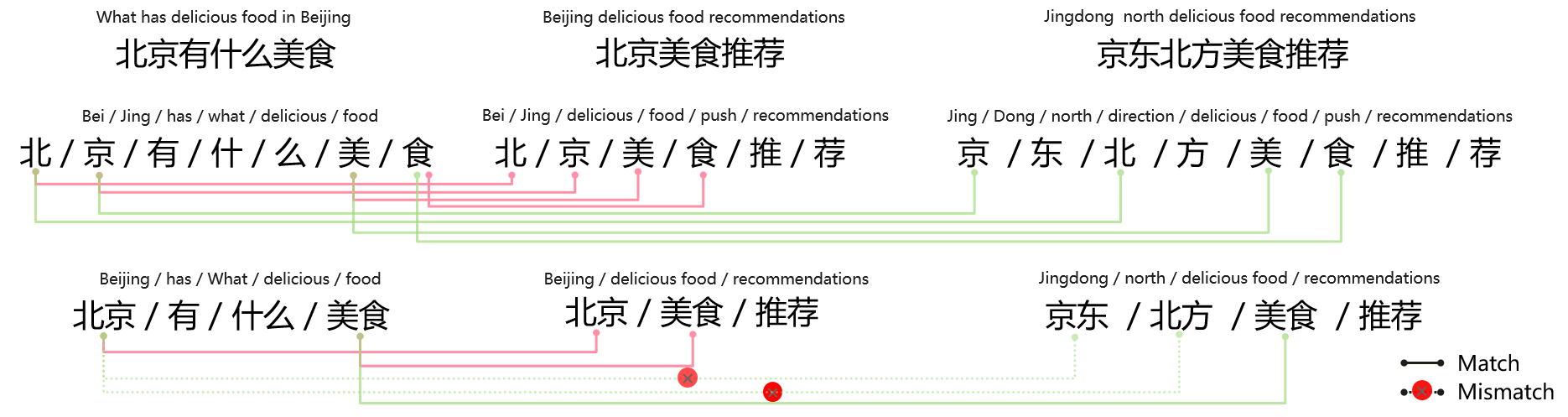}
  \caption{A example for lexicon-based retrieval. The three columns comprise the query, passage 1, and passage 2. The three rows illustrate the original text, the term-level matching results(terms derived from the tokenizer), and the word-level matching results(words generated by word segmentation), respectively.}
  \label{fig:introduction}
\end{figure}

The hybrid retrieval frameworks typically introduce a lexicon-based retrieval branch into the existing dense-vector model~\cite{gao2021coil,shen2023unifier,bge-m3}. The final matching score is computed as the sum of scores from both branches: the dense branch calculates similarity via the inner product of query and passage embeddings, while the lexicon-based branch is derived by multiplying the weights of the tokenizer-defined terms shared between the query and passage, followed by summing the resulting products. While this paradigm works adequately for English retrieval, it faces critical challenges in Chinese scenarios due to the absence of word boundaries (spaces). Specifically, lexicon-based matching relies on tokenizer-defined terms, which often fail to capture semantic nuances in Chinese. For instance, as illustrated in Fig.~\ref{fig:introduction}, term-level matching may incorrectly assign an identical score between Passage 1 and Passage 2, exposing semantic inconsistencies between term granularity and actual word meanings. This highlights the need for dedicated optimization of hybrid-based paradigms for Chinese.

It has been proven and widely accepted that in traditional lexicon-based retrieval, word-level matching properly can significantly improve performance. As illustrated in Fig.~\ref{fig:introduction} (Row 3), such improvements rely heavily on word segmentation modules to identify meaningful words. Widely adopted tools like Jieba\footnote{https://github.com/fxsjy/jieba} implement this through frequency-based heuristics, yet their lack of semantic awareness inevitably limits matching accuracy. To address this limitation, neural methods for word segmentation have emerged, utilizing bert-like models to capture semantic context~\cite{tian-etal-2020-improving-chinese,huang-etal-2020-joint-multiple,maimaiti-etal-2021-segment}. Nevertheless, these approaches typically employ two separate models for words segmentation and lexicon-based retrieval, which lacks an end-to-end optimization solution and leaves room for performance enhancement.

In this paper, we present an innovative method called HyReC, which offers an end-to-end optimization solution for hybrid-based retrieval systems in Chinese scenarios. Specifically, HyReC integrates dense-vector retrieval, lexicon-based retrieval, and the semantic union of terms into a single model. The word segmentation is defined as the semantic union of terms to distinguish the difference between tokenizer-defined terms and model-defined semantic words. Within HyReC, the $[CLS]$ embedding is utilized for dense-vector retrieval, while embeddings from other tokens are employed for sparse retrieval and the semantic union of terms. During training, we have developed a labelling tool for training the semantic union, while the dense-vector and lexicon-based retrieval components are trained using a contrastive learning approach. Once trained, HyReC conducts large-scale retrieval either through its lexicon representation using an efficient inverted index or by leveraging dense vectors with parallelizable dot-product operations. In particular, each dimension of the lexicon representation corresponds to a term in the vocabulary, with its value reflecting the importance of that term within the passage. This vocabulary includes the result from the tokenizer's definition and the newly generated words generated by the semantic union of existing terms.

Moreover, we introduce an innovative module named the Global-Local-Aware Encoder (GLAE) to facilitate consistent semantic sharing, while simultaneously minimizing the interference between the two retrieval paradigms. Since the dense-vector paradigm is designed to learn sequence-level dense representations, the lexicon-based paradigm focuses on obtaining word-level lexicon representations~\cite{shen2023unifier}. Additionally, we introduce a Normalization Module (NM) designed to align the two retrieval paradigms more reasonably, fostering mutual benefits. We normalize the matching scores of both paradigms to a 0-1 scale, rather than imposing rigid weights to enforce alignment~\cite{bge-m3}.

Our main contributions of this paper are summarized as follows: 


  $\bullet$  We propose HyReC, a novel hybrid retrieval framework tailored for Chinese scenarios that unifies dense-vector retrieval, lexicon-based retrieval, and semantic union of terms within a single model.
  
  $\bullet$ We additionally develop two key components: GLAE enables consistent semantic sharing while reducing paradigm interference, and NM achieves better alignment between two retrieval paradigms.
  
  $\bullet$ Extensive experiments demonstrate that HyReC consistently outperformed baseline algorithms on the C-MTEB retrieval benchmark, validating its effectiveness.

\begin{figure*}[t]
\centering
  \includegraphics[width=\textwidth]{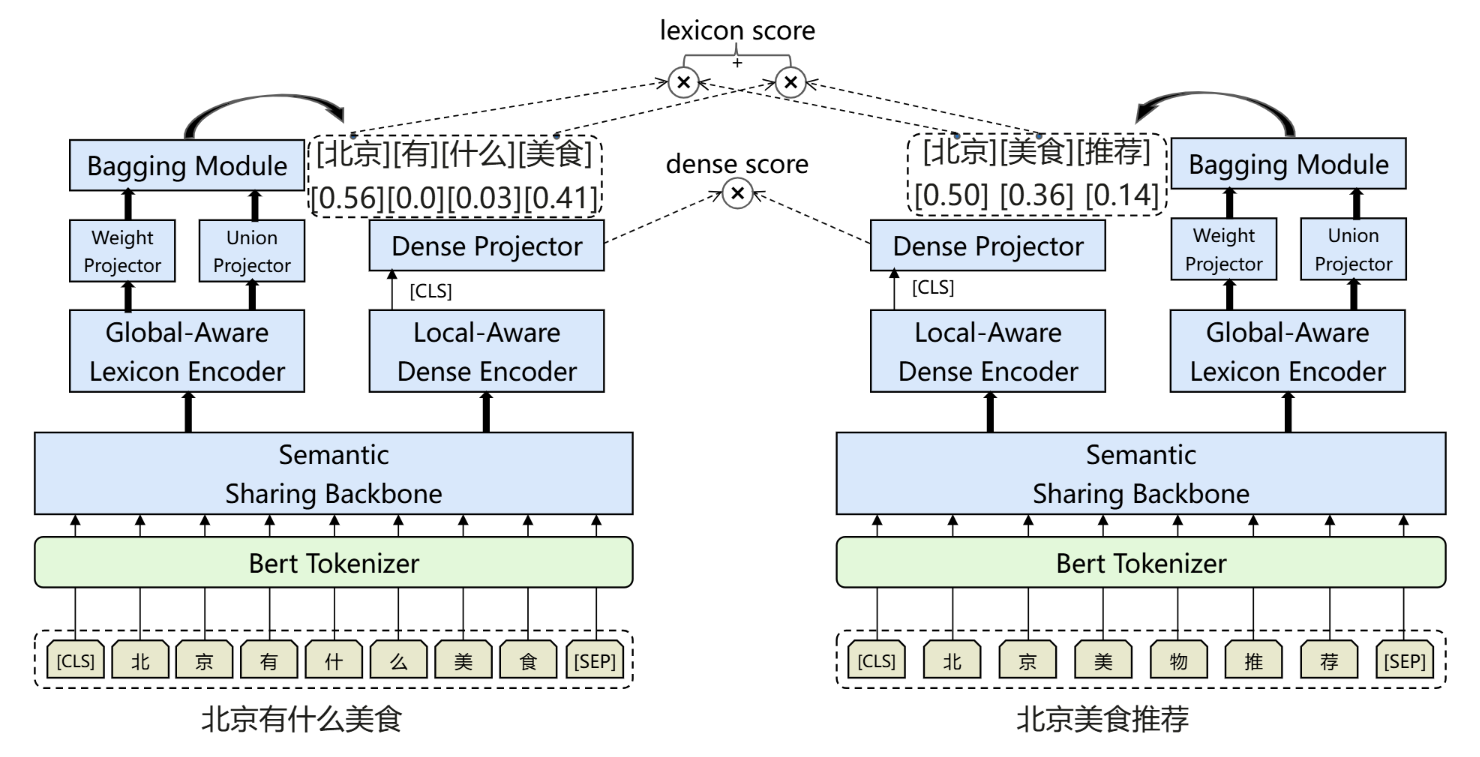}
  \caption{The architecture of our HyReC. HyReC first utilizes a semantic sharing backbone to extract low-level textual features for both paradigms. It then comprises two branches, each dedicated to learning global-aware and local-aware representations for the dense-vector retriever and the lexicon-based retriever, respectively. Additionally, a bagging module is employed to aggregate the weights and semantic union of terms, further enhancing the lexicon-based retriever's capabilities.}
  \label{fig:method_1}
\end{figure*}

\section{Related Work}
\subsection{Dense-vector Retriever}
To improve the retriever's performance, contemporary methods often focus on strategies such as selecting difficult negatives, leveraging pre-training and developing more elegant training recipes. For instance, ANCE~\cite{xiong2020approximate} introduced an innovative learning mechanism that globally selects difficult negatives from the entire corpus, utilizing an asynchronously updated approximate nearest neighbour (ANN) index. In contrast, ADORE~\cite{zhan2021optimizing} employed dynamic sampling to adaptively adjust hard negative training samples during the model training process. Moreover, Condenser~\cite{gao2021condenser} and coCondenser~\cite{gao2021unsupervised} developed a pre-training strategy specifically designed for ad-hoc retrieval to enhance the performance of the model. Recently, Bge embedding~\cite{bge_embedding}, and Jina embedding~\cite{sturua2024jinaembeddingsv3multilingualembeddingstask} have introduced a three-stage training recipe and scaled training data to further enhance the retriever's effectiveness.

\subsection{Lexicon-based Retriever}
In recent years, researchers have been fervently working to enhance context representation in the lexicon-based retriever. DeepCT~\cite{dai2019context} translates contextual term representations from BERT into term weights, deriving matching scores by multiplying the weights of terms shared between the query and passage and summing the resulting products. Similarly, SparTerm~\cite{bai2020sparterm} introduced a contextual importance predictor that accurately assesses the significance of each term within the vocabulary. Building on contextual term representations, SPLADE~\cite{formal2021splade} further introduced an innovative log-saturation effect that effectively regulates term dominance, promoting natural sparsity in the resulting representations. Additionally, TILDE~\cite{zhuang2021tilde} proposed a more efficient framework for lexicon-based retrieval by incorporating a query likelihood component.
\subsection{Hybrid-based Paradigms Retriever}
The hybrid-based paradigm has garnered significant attention from the industry owing to its superior performance. COIL~\cite{gao2021coil} used word-bag match and relied on $[CLS]$ vectors for computing relevance scores to assess hybrid-based retrievers. Bge M3~\cite{bge-m3} expanded the ability of the hybrid-based retrieval by improving the training recipe and scaled training data. The authors in~\cite{2021BERT,2024PromptReps} explored normalization for combining dense and sparse retrievals. However, our method integrates normalization module during the training phase to jointly optimize the interaction between dense and sparse retrievals, rather than only applying it during inference. Unifier~\cite{shen2023unifier} integrates dense-vector and lexicon-based retrieval into a single model with dual representing capabilities. It also introduces a self-regularization method based on list-wise agreements from these dual views. However, to improve performance in lexicon-based retrieval, Unifier replaces the embedding of the $[CLS]$ token in the lexicon encoder with that from the local-aware dense encoder. This decision increases the inference time for lexicon-based retrieval and couples the lexicon-based retrieval with dense-vector retrieval, limiting flexibility in their applications. Additionally, the previously mentioned methods place limited emphasis on this scheme within the Chinese context and ignore the union between adjacent terms.

\section{Methodology}
HyReC seamlessly integrates dense-vector retrieval, lexicon-based retrieval, and the semantic union of terms into a single model. Additionally, it incorporates GLAE to ensure consistent semantic sharing while effectively minimizing interference between the two retrieval paradigms. Ultimately, it presents NM to align these two retrieval approaches.
\subsection{Network Architecture}
As illustrated in Fig.~\ref{fig:method_1}, HyReC primarily consists of the semantic sharing backbone (detailed in Sec.\ref{ssb}), the global-aware lexicon encoder (detailed in Sec.\ref{gle}), the local-aware dense encoder (detailed in Sec.\ref{lde}), three projectors (detailed in Sec.\ref{hr}) and the bagging module (detailed in Sec.\ref{hr}). The semantic sharing backbone, the global-aware lexicon encoder and the local-aware dense encoder are collectively referred to as GLAE.

\subsubsection{Semantic Sharing Backbone}\label{ssb}
We begin by employing a semantic sharing backbone to extract low-level textual features for both retrieval paradigms, ensuring consistent semantic sharing. While the two paradigms concentrate on different levels of representation granularity( dense-vector retrieval focusing on sequence-level dense representation and lexicon-based retrieval emphasizing word-level contextualization embeddings), both paradigms delve into the semantic information of each term within the sentence. This shared exploration enables them to develop a cohesive understanding of semantic and syntactic knowledge directed toward the same retrieval targets. Like~\cite{shen2023unifier}, we also leverage a multi-layer Transformer encoder to produce the semantic sharing backbone, i.e.,

\begin{equation}
  S^{(x)} = TF\text{-}Enc([CLS]x[SEP];\theta^{(ssb)})
\end{equation}
where $TF\text{-}Enc$ refers to a multi-layer Transformer encoder that utilizes parameters $\theta^{(ssb)}$. $[CLS]$ and $[SEP]$ are special tokens by following PLMs~\cite{2019BERT}. $x$ represents either a query or a document.

\subsubsection{Global-aware Lexicon Encoder}\label{gle}
Building on the low-level textual features, we propose a representation module that generates a word-level lexicon representation. This module not only ensures consistent semantic sharing but also minimizes the interference from sequence-level dense representation associated with the dense retrieval paradigm. Unlike the approach taken in~\cite{shen2023unifier}, we refrain from replacing the embedding of the $[CLS]$ token with that from the local-aware dense encoder for two key reasons: first, to reduce the inference time of the lexicon-based retrieval; and second, to decouple the lexicon-based retrieval and the dense-vector retrieval, allowing for more flexibility in their application. Given that the word-level lexicon representation captures global vocabulary space information, we designate this module as the global-aware lexicon encoder. To achieve this, we utilize an additional multi-layer Transformer encoder to process $S^{(x)}$. This can be expressed as

\begin{equation}
  L^{(x)} = TF\text{-}Enc(S^{(x)};\theta^{(gle)})
\end{equation}

where this module is parameterized by $\theta^{(gle)}$, which is distinct from $\theta^{(ssb)}$, the resulting $L^{(x)}$ denotes a word-level lexicon representation of the input text $x$, which is utilized for lexicon-based retrieval.

\subsubsection{Local-aware Dense Encoder}\label{lde}
Additionally, building on the low-level textual features, we present another representation module that generates a sequence-level dense representation. This module not only ensures consistent semantic sharing but also reduces interference from the word-level lexicon representation associated with the lexicon retrieval paradigm. Since the sequence-level dense representation does not incorporate global vocabulary space information, which captures local contextualization, we refer to this module as the local-aware dense encoder. To achieve this, we apply another multi-layer Transformer encoder to $S^{(x)}$, This can be written as
\begin{equation}
  D^{(x)} = TF\text{-}Enc(S^{(x)};\theta^{(lde)})
\end{equation}
where this module is parameterized $\theta^{(lde)}$ , the resulting $D^{(x)}$ denotes a sequence-level dense representation of the input text $x$, which is employed for dense-vector retrieval. The dimension of  $S^(x), L^(x)$ and $D^(x)$ is $[B, N, H]$, where B represents the batch size, N denotes the input sequence length, and H indicates the hidden size of the model.

\subsubsection{Hybrid Retrieval}\label{hr}
After obtaining the word-level lexicon representation $L^{(x)}$ and sequence-level dense representation $D^{(x)}$ of the input text $x$, we employ three projectors( i.e., weight projector, union projector and dense projector) to acquire the term weight, term union and dense vector, respectively. 

To achieve term union, the union projector combines the word-level lexicon representation  $L^{(x)}$ into four classification probabilities, indicating `S' (single term word) `B' (the beginning position of the word), `M' (the middle position of the word), and `E' (the ending position of the word), respectively. This is expressed as:
\begin{equation}
U_{term_i} = softmax(w_ul_i + b_u)
\end{equation}
where ${term_i}$ is the $ith$ term or token in input $x$. $w_u$ and $b_u$ are linear weights and bias of the union projector module, respectively. $l_i$ is $ith$ token's word-level lexicon representations from $L^{(x)}$.

For term weight, we adopt a method inspired by the recent TILDEv2~\cite{zhuang2021fast}, which optimizes memory usage by storing only the scores of tokens that appear in current passages rather than the entire vocabulary. Differing from the original TILDEv2, our lexicon-based retrieval incorporates term union information to enhance performance. The weight projector  integrates the word-level lexicon representation $L^{(x)}$ to produce a term importance score:
\begin{equation}
W_{term_i} = log(1 + ReLU(w_wl_i + b_w))
\end{equation}
where $w_w$ and $b_w$ are linear weights and bias of the weight projector module, respectively.

Lastly, the dense projector combines the representation of special token $[CLS]$ from the sequence-level dense representation $D^{(x)}$ to generate a sequence-level dense vector, which is utilized for dense-vector retrieval:
\begin{equation}
D_{vec} = (w_dd_{CLS} + b_d)
\end{equation}
where $w_d$ and $b_d$ are linear weights and bias of the dense projector module, respectively. $d_{CLS}$ is $[CLS]$ representations from $D^{(x)}$.

In the inference, we utilize the Bagging module to aggregate both the weights and the semantic unions of terms. To elaborate, we proceed as follows: first, based on $U_{term_i}$, we derive the result $U_{word_j}$ for the $jth$ semantic union. Notably, $U_{word_j}$ may encompass multiple terms or tokens when $U_{term_i}$ belongs to the set $\{B, M, E\}$; Second, we compute the weight $W_{word_j}$ associated with $U_{word_j}$. as follows:
\begin{equation}
U_{word_j} = max(U_{term_i});  term_i\in word_j
\end{equation}
\begin{equation}
W_{word_j} = max(W_{term_i});  term_i\in word_j
\end{equation}
where $word_j$ represents a word that consists of more than one term or token.

The final matching score for hybrid retrieval is calculated as the sum of the matching scores from both the lexicon retrieval and the dense retrieval, i.e., 
\begin{equation}
S(q, p) = S^{lex}(q, p) + S^{den}(q, p)
\end{equation}
where $S^{lex}(q, p)$ and $S^{den}(q, p)$ denote the matching score of the lexicon retrieval and the dense retrieval, respectively. The matching score of the lexicon-based method is calculated by weights multiplications of the common terms shared in the query and the passages. 

\begin{equation}
S^{lex}(q, p) = \sum_{i, j} W^q_{\hat{term_i}} W^p_{\hat{term_j}}
\end{equation}
where $W^q_{\hat{term_i}}$, $W^p_{\hat{term_j}}$ represents the weight of the $ith$ term( or word) from the query and the passage, respectively. $\hat{term_i}$ is derived from both $term_i$ and $word_i$( similarly for $\hat{term_j}$.).



\begin{equation}
S^{den}(q, p) = D^q_{vec} \cdot D^p_{vec}
\end{equation}
where $D^q_{vec}$ and $D^p_{vec}$ are sequence-level dense vectors of query and passage, respectively.
\subsection{Loss}
Given a query $q$ and a set of $n$ passages $D = \{ p^+, \hat{p}_1, \hat{p}_2, ..., \hat{p}_{n-1}\}$. The lexicon retrieval and the dense retrieval task are acquired by the ranking objective with a contrastive loss. Thus, their training loss is
\begin{equation}
\mathcal{L}_{*} = -log\frac{e^{S^*(q, p^+)/\tau}}{e^{S^*(q, p^+))/\tau}+\sum_{\hat{P}} e^{S^*(q, \hat{p}_j)/\tau}}
\end{equation}
where $\tau$ is the temperature parameter. $*$ is $lex$ or $den$, which denotes the matching score of the lexicon retrieval and the dense retrieval, respectively. $p$ and $q^+$ represent the paired texts, $\hat{p}_j \in \hat{P} $ denotes a hard negative.

The semantic union loss, represented as $\mathcal{L}_{union}$, is employed by the cross-entropy loss, as represented below.
 
\begin{equation}
\mathcal{L}_{union} = -\sum_i y_i log(U_{term_i})
\end{equation}
where $y_i$ represents the labels of the term union (for more details, see Section~\ref{lablling_tool}).

The total loss for our HyReC is
\begin{equation}
\mathcal{L} = \mathcal{L}_{lex} + \mathcal{L}_{den} + \mathcal{L}_{union}
\end{equation}

\subsection{Normalization Module (NM)}
In lexicon-based retrieval, matching scores are computed through weighted summation of identical terms, where the score range varies significantly depending on term weights. Similarly, dense retrieval scores obtained through vector dot products also exhibit unpredictable ranges. This discrepancy in score distributions makes direct combination problematic, necessitating normalization to align their scales. We first perform L2 normalization on the sequence-level dense representation $D^{(x)}$ before computing dot products. For the lexicon-based branch, we employ an attention mask to identify valid tokens, followed by L2 normalization of the term importance score vectors $W_{term_i}$ for these selected tokens. The benefits of this module are as follows: First, it mitigates training instability caused by score disparities, which otherwise lead to inconsistent model preferences for dense or sparse retrieval across samples. Second, the normalization process effectively balances the contribution of each branch, allowing for more stable optimization. Finally, by aligning the score distributions, the model can learn more meaningful combination weights during training.


\section{Experiments}

\subsection{Implementation Details}
We follow the training recipe of BGE~\cite{bge_embedding} to ensure a fair comparison. See details of method and datasets in Appendix~\ref{Training_Recipe}.

\begin{table*}
  \centering
  \begin{tabular}{llllllllll}
    \hline
    Model & T2 & MM & Du & Covid & Cmed & Ecom & Med & Video & Avg \\
    \hline
luotuo-bert-medium & 58.67 & 55.31 & 59.36 & 55.48 & 18.04 & 40.48 & 29.8 & 38.04 & 44.4 \\
text2vec-base-chinese & 51.67 & 44.06 & 52.23 & 44.81 & 15.91 & 34.59 & 27.56 & 39.52 & 38.79 \\
m3e-base & 73.14 & 65.45 & 75.76 & 66.42 & 30.33 & 50.27 & 42.8 & 51.11 & 56.91 \\
OpenAI & 69.14 & 69.86 & 71.17 & 57.21 & 22.36 & 44.49 & 37.92 & 43.85 & 52.0 \\
multilingual-e5-base & 70.86 & 76.04 & 81.64 & 73.45 & 27.2 & 54.17 & 48.35 & 61.3 & 61.63 \\
BAAI/bge-base-zh & 83.35 & \textbf{79.11} & 86.02 & 72.07 & \textbf{41.77} & 63.53 & 56.64 & \textbf{73.76} & 69.53 \\
BGE-m3-sparse & 71.80 & 59.31 & 71.53 & 76.57 & 24.32 & 50.76 & 43.78 & 58.68 & 57.08 \\
BGE-m3-dense & 81.07 & 77.25 & 84.03 & 76.56 & 33.78 & 58.39 & 54.27 & 56.95 & 65.29 \\
BGE-m3-hybrid & 83.04 & 77.57 & 84.52 & 79.22 & 33.28 & 60.65 & 55.35 & 63.13 & 67.10 \\
HyReC-base-sparse & 73.00 & 69.43 & 74.58 & 74.41 & 30.76 & 59.64 & 48.29 & 67.07 & 62.15 \\
HyReC-base-dense & 82.93 & 77.40 & \textbf{89.13} & 76.06 & 34.42 & 61.31 & 57.30 & 72.16 & 68.84 \\
HyReC-base-hybrid & \textbf{84.03} & 77.81 & 87.31 & \textbf{79.53} & 38.47 & \textbf{64.82} & \textbf{58.89} & 73.46 &  \textbf{70.54} \\
\hline
multilingual-e5-small & 71.39 & 73.17 & 81.35 & 72.82 & 24.38 & 53.56 & 44.84 & 58.09 & 59.95 \\
BAAI/bge-small-zh & 77.59 & 67.56 & 77.89 & 68.95 & \textbf{35.18} & 58.17 & 49.9 & 69.33 & 63.07 \\
HyReC-small-sparse & 73.20 & 68.23 & 74.67 & 75.93 & 28.67 & 59.30 & 46.64 & 68.48 & 61.89 \\
HyReC-small-dense & 76.90 & 70.30 & 82.76 & 72.43 & 33.81 & 55.04 & 51.02 & 66.05 & 63.54 \\
HyReC-small-hybrid & \textbf{80.52} & \textbf{73.29} & \textbf{82.82} & \textbf{76.47} & 34.93 & \textbf{61.43} & \textbf{53.27} & \textbf{71.14} & \textbf{66.73} \\
    \hline
  \end{tabular}
  \caption{\label{State-of-the-art}
    The experimental results on the C-MTEB retrieval benchmark are evaluated using $nDCG@10$. T2, MM, Du, Covid, Cmed, Ecom, Med and Video correspond to the T2Retrieval, MMarcoRetrieval, DuRetrieval, CovidRetrieval, CmedqaRetrieval, EcomRetrieval, MedicalRetrieval and VideoRetrieval development settings in C-MTEB retrieval benchmark.
  }
\end{table*}

\subsubsection{Labelling Tool for Semantic Union}\label{lablling_tool}
We combine the Jieba frequency word segmentation module with manually crafted regular expressions to generate labels for the semantic union. Since the Jieba framework relies primarily on frequency information and exhibits limited generalization, we enhance it by incorporating regular expressions to identify additional patterns, such as numeric expressions, quantity phrases, product models, and version identifiers. Labelling the semantic union involves the following steps: 1) Extracting the segmented words from the input and identifying the offsets of these segmented words within the original string; 2) Obtaining the tokenization results and their respective offsets; 3) Producing the labelling results by aligning the offsets between the segmented words and the tokenizer outputs. The offsets of the segmented words align a single token and the corresponding token is labelled as `S' (indicating 0 class). For offsets that encompass multiple tokens, the tokens are labelled sequentially as `B' (indicating 1 class), `M' (indicating 2 class), and `E' (indicating 3 class).
\subsubsection{Evalution Metrics }
In this paper, we explore the application of a hybrid-based paradigm within Chinese retrieval scenarios, focusing primarily on Chinese retrieval experiments. We adopt the C-MTEB~\cite{bge_embedding} retrieval benchmark as our standard, given its prominence in the field. Adhering to the official benchmark protocols, we evaluate our method using Pyserini~\cite{2021Pyserini} and utilize $nDCG@10$ as the primary evaluation metric.
\subsubsection{Experimental Setups}
For pre-training with a large volume of unsupervised data, we set the batch size to 512, with a maximum query and passage length of 512 tokens. The learning rate is configured at $1 \times 10^{-4}$, the warmup ratio is 0.1, and the weight decay is set to 0.01. This pre-training process is conducted across 16 V100 (32GB) GPUs.

In the two fine-tuning stages, we utilize a batch size of 64 for the small-scale model and 30 for the base-scale model. Additionally, we set the maximum lengths for queries and passages to 64 and 256 tokens, respectively. We perform 5 epochs with a learning rate of $5 \times 10^{-5}$, a temperature parameter of 0.05, and a weight decay of 0.01. The fine-tuning stage is executed on 8 3090 (24GB) GPUs. For the high-quality fine-tuning stage, we sample 3 negative instances for each query. 

The small-scale model( with 38M parameters) is composed of a single-layer semantic sharing backbone, a single-layer global-aware lexicon encoder, and a single-layer local-aware dense encoder. In contrast, the base-scale model( with 153M parameters) incorporates a 5-layer semantic sharing backbone, a 7-layer global-aware lexicon encoder, and a 7-layer local-aware dense encoder. Notably, in small cases due to the constraints of our machine, we have opted not to scale our model to a large scale( like BAAI/bge-large-zh~\cite{bge_embedding}).

\begin{table}
\centering
\small
  \caption{Ablation study on the effectiveness of each component on C-MTEB retrieval benchmark( $SU$ means the semantic union of terms).}
  \label{tab:Effectiveness_each_component}
  \begin{tabular}{ccccccl}
   NM & GLAE & SU & Lexicon & Dense & Hybrid \\
  \toprule
               & \checkmark & \checkmark & 56.89 & 57.80 & 58.53 \\
    \checkmark &            & \checkmark & 60.80 & 60.82 & 65.41 \\
    \checkmark & \checkmark &            & 60.65 & 63.13 & 66.14 \\
    \checkmark & \checkmark & \checkmark & \textbf{61.89} & \textbf{63.54} & \textbf{66.73} \\
  \bottomrule
\end{tabular}
\end{table}

\subsection{Main Evaluation}
We conducted experiments using the C-MTEB retrieval benchmark~\cite{bge_embedding} to compare HyReC with the existing methods listed in Tab.~\ref{State-of-the-art}. Our HyReC model exhibits remarkable advancements on the C-MTEB retrieval benchmark, It significantly outperforms Bge~\cite{bge_embedding} on $nDCG@10$ with a margin of +3.66\%. A comparable improvement was observed when we scaled our model to a base size, resulting in an additional gain of +1.01\% (notably, without the constraints of our machine, this improvement would be even more pronounced). Furthermore, the integration of lexicon-based and dense-vector retrieval leads to notable enhancements in retrieval performance, with improvements of +4.84\% for lexicon-based retrieval and +3.19\% for dense-vector retrieval. Ultimately, compared to the BGE-m3-hybrid approach which employs the same method~\cite{bge-m3}, our hybrid-based retrieval method achieved a significant improvement in retrieval performance, reaching a 3.44\% enhancement, demonstrating its outstanding effectiveness. Moreover, even without the inclusion of CovidRetrieval in the training data, our approach showcases its remarkable ability to generalize across diverse datasets.


\subsection{Ablation Studies}

\subsubsection{Effectiveness of Each Component}

The contributions of different components of HyReC are listed in Tab.~\ref{tab:Effectiveness_each_component}. Utilizing the small-scale model of HyReC for this ablation study, we observed that the removal of the NM module leads to a significant drop in performance, highlighting the detrimental effects of an unpredictable score range on unstable training. The GLAE module was configured by adjusting the number of layers in the semantic sharing backbone, global-aware lexicon encoder, and local-aware dense encoder. The removal of the GLAE module entailed eliminating both the global-aware lexicon encoder and the local-aware dense encoder, while setting the number of layers in the semantic sharing backbone to two. The removal of the GLAE module leads to a decline in both lexicon-based (61.89\% to 60.80\%) and dense-vector (63.54\% to 60.82\%) retrieval performance, highlighting the interplay between the two retrieval paradigms. Furthermore, we conducted two additional experiments: 1) training exclusively on the lexicon-based retrieval task, which yielded an $nDCG@10$ score of 53.05\%, and 2) training solely on the dense-vector retrieval task, resulting in an $nDCG@10$ score of 63.11\%. Consequently, the semantic sharing backbone promotes consistent semantic sharing, as evidenced by substantial improvements—rising from 53.05\% to 61.89\% for lexicon-based retrieval and from 63.11\% to 63.54\% for dense-vector retrieval, with the most notable enhancement observed in lexicon-based retrieval. Additionally, introducing semantic union results in marked improvements: lexicon-based (rising from 60.65\% to 61.89\%), dense-vector (increasing from 63.13\% to 63.54\%), and hybrid-based retrieval (growing from 66.14\% to 66.73\%). This clearly illustrates that semantic union not only enhances lexicon-based retrieval but also positively affects dense-vector retrieval. The ablation studies presented in Tab.~\ref{tab:Effectiveness_each_component} verify the effectiveness of each module in our HyReC.

\begin{figure}[t]
  \includegraphics[width=0.8\columnwidth]{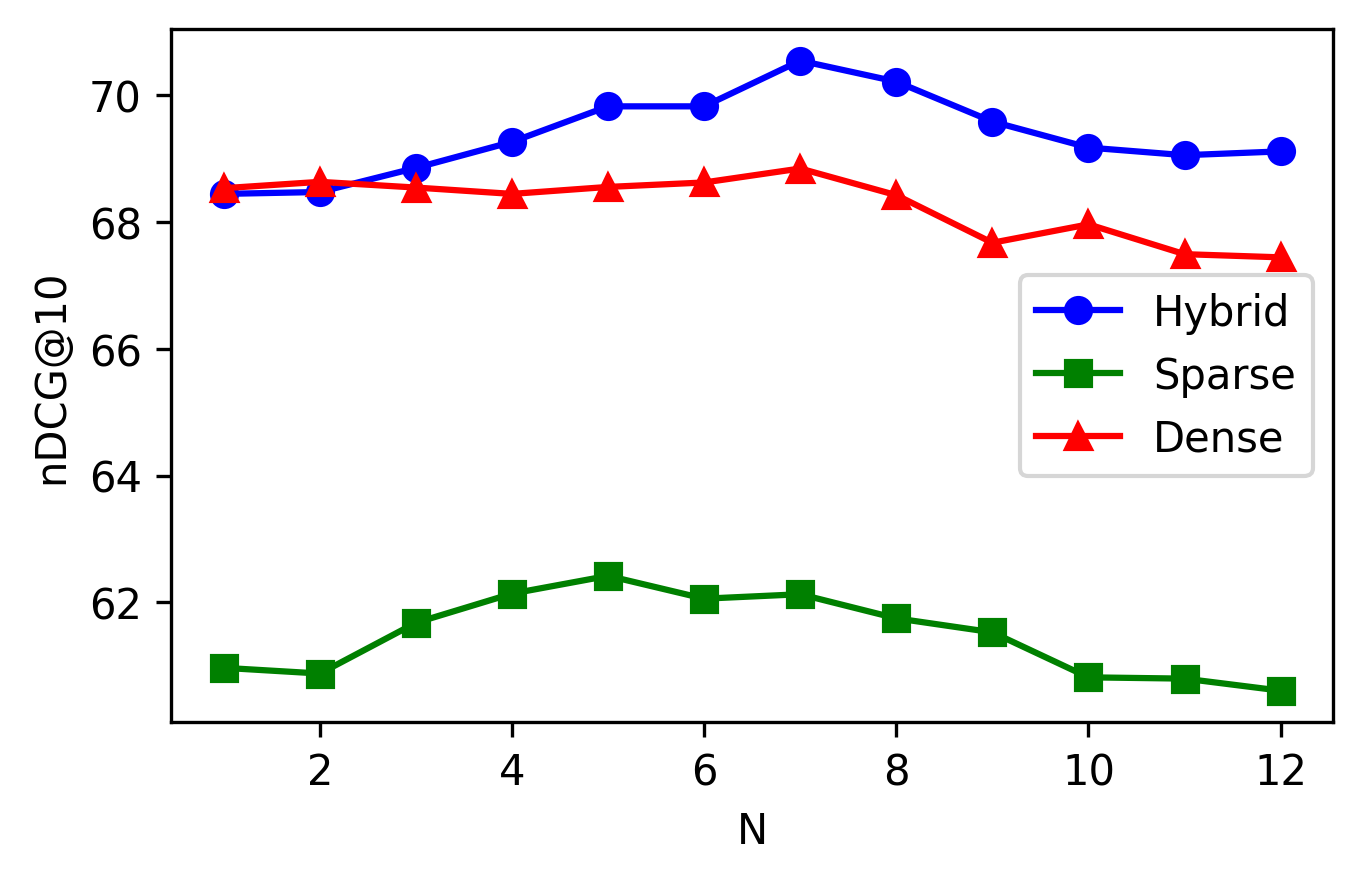}
  \caption{Ablation study on the performance of GLAE by the base-scale model, where the row axis $N$ is the layer number of the global-aware lexicon encoder or the local-aware dense encoder and the layer number of the semantic sharing backbone is $12-N$.}
  \label{fig:GLAE}
\end{figure}

\subsubsection{Parameters of GLAE}
As depicted in Fig.~\ref{fig:GLAE}, increasing the number of layers in the global-aware lexicon encoder or the local-aware dense encoder(while progressively reducing the layer number of the semantic sharing backbone) initially leads to an increase in all 
 $nDCG@10$ scores. This initial enhancement can be attributed to a decrease in the interplay between the two retrieval paradigms. Beyond a certain point, the scores begin to decline, which is a consequence of reduced consistent semantic sharing. 

\subsection{Ablation of the Semantic Union of Terms}
We conducted experiments to validate the proposed semantic union of terms. As illustrated in Tab. ~\ref{tab:Effectiveness_sut}, the $nDCG@10$ performance of the semantic union surpasses the frequency-based method by 4.19\%, 0.41\% and 0.52\% in lexicon-based, dense-vector and hybrid-based retrieval performance, respectively. These results clearly show that our proposed semantic union of terms outperforms the frequency-based method, i.e., Jieba. Additionally, we visualize the outputs of both the semantic union of terms and Jieba to assess the semantic information utilized in generating segmented words. As demonstrated in Tab. ~\ref{tab:Effectiveness_sut_vis}, the proposed semantic union of terms effectively comprehends the semantic nuances of the terms and achieves meaningful word segmentation by this semantic understanding.
\begin{table}
\centering
\small
  \caption{Ablation study on the performance of the semantic union of terms on C-MTEB retrieval benchmark.}
  \label{tab:Effectiveness_sut}
  \begin{tabular}{ccccl}
  & Method & Lexicon & Dense & Hybrid \\
  \toprule
   & Jieba & 57.70 & 63.13 & 66.21 \\
   & HyReC & \textbf{61.89} & \textbf{63.54} & \textbf{66.73} \\
  \bottomrule
\end{tabular}
\end{table}

\begin{CJK}{UTF8}{gbsn}

\begin{table}
\centering
\scriptsize
  \caption{The case study of the semantic union. Red words are incorrect segmentation.}
  \label{tab:Effectiveness_sut_vis}
  \begin{tabular}{|c|c|}
  \hline
  Sentence & 李一一一下子想不起她是谁 \\
  & LiYiyi immediately couldn't remember who she was \\
  \hline
   Jieba &  \textcolor{red}{李}/ \textcolor{red}{一一}/ 一下子/ 想不起/ 她/ 是/ 谁\\
   & \textcolor{red}{Li}/ \textcolor{red}{Yiyi}/ immediately/ couldn't remember/ she/ was/ who\\
   \hline
   HyReC &  李一一/ 一下子/ 想不起/ 她/ 是/ 谁\\
    & LiYiyi/ immediately/ couldn't remember/ she/ was/ who\\
   \hline
   Sentence & 你告诉我光弱一端 \\
  & You tell me weak light side \\
  \hline
   Jieba &  你/ 告诉/ \textcolor{red}{我光弱}/ 一端\\
   &  You/ tell/ \textcolor{red}{my weak light}/ side \\
   \hline
   HyReC &  你/ 告诉/ 我/ 光弱/ 一端\\
   & You/ tell/ me/ weak light/ side \\
  \hline
\end{tabular}
\end{table}
\end{CJK}

\section{Conclusion}
In this paper, we introduced HyReC, an innovative end-to-end optimization method specifically designed for hybrid-based retrieval systems in the Chinese context. HyReC effectively integrates dense-vector retrieval, lexicon-based retrieval, and the semantic union of terms into a cohesive model to enhance overall performance. Additionally, our method incorporates two pivotal modules: (1) the Global-Local-Aware Encoder (GLAE), which facilitates consistent semantic sharing while minimizing interference between the retrieval paradigms, and (2) the Normalization Module (NM), which further fine-tunes the alignment between these retrieval paradigms. Our experimental results reveal that HyReC significantly outperforms the baseline, achieving remarkable improvements in $nDCG@10$ (+3.66\% for the small-scale model and +1.01\% for the base-scale model). The evaluations conducted on the C-MTEB retrieval benchmark conclusively demonstrate the effectiveness of our proposed approach.

\section{Limitations}
While our study presents a novel optimization module, semantic union of terms, tailored for enhancing retrieval tasks in Chinese retrieval scenarios, it is important to acknowledge two key limitations. First, the proposed module is specifically designed and optimized for Chinese language expressions, and its applicability to other languages remains unexplored. This limitation arises from the inherent linguistic characteristics embedded in the semantic union of terms, which are currently aligned with Chinese and may not directly generalize to multilingual contexts. Future work could investigate the adaptation of this module to other languages by incorporating cross-lingual semantic representations. Second, the experimental validation of our approach is confined to retrieval tasks on the C-MTEB benchmark, and its performance in other tasks, such as classification or clustering, has not been evaluated. This restriction stems from the fact that the semantic union of terms is inherently optimized for retrieval matching, and its effectiveness in broader applications remains an open question. Extending the evaluation to additional domains could provide a more comprehensive understanding of the module's versatility and potential impact. Additionally, due to the constraints of our GPU resources, we were unable to scale the model to a larger size(like BAAI/bge-large-zh~\cite{bge_embedding}) and the optimization of the base-scale model may not have been fully realized, limiting its potential performance.




\clearpage
\appendix
\begin{table*}[t]
  \centering
  \begin{tabular}{llllllllll}
    \hline
    Model & T2 & MM & Du & Covid & Cmed & Ecom & Med & Video & Avg \\
    \hline
BAAI/bge-base-zh & 82.38 & \textbf{88.73} & 87.27 & 86.78 & \textbf{51.39} & \textbf{80.20} & 65.90 & 86.00 & 78.58 \\
BGE-m3-sparse & 70.19 & 71.04 & 72.19 & 86.56 & 29.89 & 65.00 & 51.10 & 74.20 & 65.02 \\
BGE-m3-dense & 80.12 & 88.18 & 86.25 & 88.83 & 42.33 & 74.60 & 62.80 & 72.30 & 74.43 \\
BGE-m3-hybrid & 81.68 & 88.35 & 86.52 & 90.25 & 41.31 & 76.40 & 63.60 & 78.50 & 75.83 \\
HyReC-base-sparse & 72.28 & 80.62 & 76.06 & 87.20 & 37.10 & 74.40 & 55.70 & 82.90 & 70.78 \\
HyReC-base-dense & 81.70 & 87.93 & \textbf{89.55} & 88.41 & 42.59 & 76.30 & 66.30 & 85.40 & 77.27 \\
HyReC-base-hybrid & \textbf{82.73} & 87.83 & 88.28 & \textbf{90.99} & 46.83 & 79.40 & \textbf{68.10} & \textbf{86.90} & \textbf{78.88} \\
\hline
BAAI/bge-small-zh & 76.29 & 79.22 & 79.80 & 82.35 & \textbf{44.08} & 73.30 & 58.10 & 82.40 & 71.94 \\
HyReC-small-sparse & 72.45 & 79.77 & 75.74 & 88.09 & 34.91 & 72.80 & 54.60 & 83.80 & 70.27 \\
HyReC-small-dense & 75.94 & 81.45 & 83.69 & 84.56 & 42.60 & 70.60 & 60.40 & 81.00 & 72.53 \\
HyReC-small-hybrid & \textbf{79.33} & \textbf{84.20} & \textbf{84.05} & \textbf{88.25} & 42.43 & \textbf{76.30} & \textbf{62.10} & \textbf{86.10} & \textbf{75.34} \\
    \hline
  \end{tabular}
  \caption{\label{State-of-the-art_1}
    The experimental results on the C-MTEB retrieval benchmark are evaluated using $Recall@10$. T2, MM, Du, Covid, Cmed, Ecom, Med and Video correspond to the T2Retrieval, MMarcoRetrieval, DuRetrieval, CovidRetrieval, CmedqaRetrieval, EcomRetrieval, MedicalRetrieval and VideoRetrieval development settings in C-MTEB retrieval benchmark.}
  
\end{table*}

\begin{table*}[t]
  \centering
  \begin{tabular}{llllllllll}
    \hline
    Model & T2 & MM & Du & Covid & Cmed & Ecom & Med & Video & Avg \\
    \hline
BAAI/bge-base-zh & 92.13 & 74.52 & 90.95 & 70.77 & \textbf{44.44} & 59.32 & 53.62 & 67.85 & 69.20 \\
HyReC-base-sparse & 85.04 & 66.27 & 85.47 & 70.22 & 34.38 & 54.97 & 45.95 & 61.98 & 63.03 \\
HyReC-base-dense & 91.62 & 74.43 & \textbf{94.77} & 72.14 & 37.00 & 56.58 & 54.48 & 67.83 & 68.61 \\
HyReC-base-hybrid & \textbf{92.47} & \textbf{74.92} & 93.26 & \textbf{75.89} & 41.50 & \textbf{60.24} & \textbf{56.05} & \textbf{69.08} & \textbf{70.43} \\
\hline
BAAI/bge-small-zh & 88.47 & 64.27 & 86.59 & 64.66 & 38.04 & 53.39 & 47.32 & 65.15 & 63.49 \\
HyReC-small-sparse & 85.12 & 64.96 & 85.30 & 71.88 & 32.08 & 55.02 & 44.10 & 63.52 & 62.75 \\
HyReC-small-dense & 87.64 & 67.19 & \textbf{90.70} & 68.65 & 36.50 & 50.12 & 48.13 & 61.23 & 63.77 \\
HyReC-small-hybrid & \textbf{90.29} & \textbf{70.19} & 90.53 & \textbf{72.68} & \textbf{38.26} & \textbf{56.78} & \textbf{50.53} & \textbf{66.30} & \textbf{66.94} \\
    \hline
  \end{tabular}
  \caption{\label{State-of-the-art_2}
   The experimental results on the C-MTEB retrieval benchmark are evaluated using $MRR@10$.
  }
\end{table*}

\begin{table*}[t]
  \centering
  \begin{tabular}{llllllllll}
    \hline
    Model & T2 & MM & Du & Covid & Cmed & Ecom & Med & Video & Avg \\
    \hline
BAAI/bge-base-zh & 76.02 & 73.93 & 76.98 & 70.80 & \textbf{35.18} & 59.32 & 53.52 & 67.85 & 64.20 \\
HyReC-base-sparse & 63.18 & 65.59 & 64.48 & 70.27 & 25.84 & 54.97 & 45.95 & 61.98 & 56.53 \\
HyReC-base-dense & 75.03 & 73.79 & \textbf{82.58} & 72.05 & 29.03 & 56.58 & 54.47 & 67.88 & 63.93 \\
HyReC-base-hybrid & \textbf{76.32} & \textbf{74.39} & 80.20 & \textbf{75.82} & 32.71 & \textbf{60.24} & \textbf{55.99} & \textbf{69.08} & \textbf{65.60} \\
\hline
BAAI/bge-small-zh & 68.59 & 63.60 & 68.39 & 64.64 & 29.24 & 53.39 & 47.27 & 65.17 & 57.54 \\
HyReC-small-sparse & 63.44 & 64.27 & 64.98 & 71.99 & 23.87 & 55.02 & 44.10 & 63.52 & 56.40 \\
HyReC-small-dense & 67.76 & 66.47 & 74.37 & 68.52 & 28.06 & 50.12 & 48.08 & 61.28 & 58.08 \\
HyReC-small-hybrid & \textbf{72.01} & \textbf{69.58} &\textbf{ 74.44} & \textbf{72.65} & \textbf{29.49} & \textbf{56.78} & \textbf{50.48} & \textbf{66.30} & \textbf{61.47} \\
    \hline
  \end{tabular}
  \caption{\label{State-of-the-art_3}
    The experimental results on the C-MTEB retrieval benchmark are evaluated using $MAP@10$. 
  }
\end{table*}

\section{Training Recipe}\label{Training_Recipe}
Our training pipeline consists of both pre-training and fine-tuning phases. During fine-tuning, the global-aware lexicon encoder and local-aware dense encoder are initialized with the same pre-trained parameters. Specifically, we partition the pre-trained BERT model into two components: (1) the semantic-sharing backbone and (2) the encoder module. The latter is then used to initialize both the global and local encoders. 

$\bullet$ \textbf{Pre-Training.} Utilizing the RetroMAE~\cite{xiao2022retromaepretrainingretrievalorientedlanguage} method, a variant of mask language modeling, we leverage the Wudao~\cite{2021WuDaoCorpora} corpora to pre-train our model, which means we do not use any pre-trained language models.

$\bullet$ \textbf{Preliminary Fine-tuning.} At this stage, we gather text pairs from various open web sources, such as Zhihu and Baike. To enhance the quality of our dataset, we employ a third-party model, Text2Vec-Chinese\footnote{https://huggingface.co/GanymedeNil}, to filter out noisy data by applying a threshold of 0.43. Through this process, we successfully filter 160 million text pairs from the unlabeled corpora. Finally, the pre-trained model undergoes fine-tuning on this carefully curated corpus, which empowers it to effectively differentiate between the paired texts. Contrastive learning is employed to achieve local-aware dense representations and global-aware lexicon representations, while classification learning is utilized for semantic union. In-batch negative samples are adopted during training for contrastive learning.

$\bullet$ \textbf{High-quality Fine-tuning.} The model undergoes additional fine-tuning using a set of high-quality text pairs, which includes $T^2$-Ranking~\cite{20081}, DURreader~\cite{he2018dureaderchinesemachinereading}, mMARCO~\cite{bonifacio2022mmarcomultilingualversionms}, CMedQA-v2~\cite{2018Multi} and multi-cpr~\cite{2022Multi}. In total, there are 118,944,5 paired texts, most of which are curated through human annotation to ensure their high quality. During this stage, both contrastive learning and classification learning are employed to further refine the model. In contrastive learning, we not only utilize in-batch negative samples but also implement an ANN-style sampling strategy~\cite{2021Approximate} to generate hard negative samples. This stage features two key distinctions from the preliminary fine-tuning: firstly, it incorporates high-quality text pairs with human annotations for training; secondly, the negative sampling process is enhanced by the inclusion of hard negative samples.

\section{ANN-style Hard Negative Mining}\label{ANN-style}
Our ANN-style hard negative mining involves:

$\bullet$ building an index using the first-stage model.

$\bullet$ retrieving the top 100 passages per query.

$\bullet$ sampling hard negatives from non-positive passages ranked 20th–100th.

\section{Evaluation in Other Metrics}
As illustrated in Tables~\ref{State-of-the-art_1}-\ref{State-of-the-art_3}, our retrieval method significantly enhances performance. This is evidenced by its impressive results across $Recall@10$, $MRR@10$, and $MAP@10$ metrics, highlighting its exceptional effectiveness( $MRR@10$ and $MAP@10$ metrics exclude the bge-m3 model due to the absence of pertinent evaluation codes on its official website.).


\end{document}